\documentclass[aps,prd,reprint,nofootinbib]{revtex4-1}

\usepackage{hyperref} \usepackage{amsmath} \usepackage{amssymb}
\usepackage{graphicx} \usepackage{color}

\newcommand{\order}[1]{\mathcal{O}\left( #1 \right)}

\newcommand{\mathset}[1]{\left\{ #1 \right\}}

\newcommand{\xmin}{x_\mathrm{min}}

\newcommand{\be}{\begin{equation}}
\newcommand{\ee}{\end{equation}}
\newcommand{\ba}{\begin{eqnarray}}
\newcommand{\ea}{\end{eqnarray}}

\DeclareMathOperator{\erf}{erf}
\DeclareMathOperator{\erfc}{erfc}

\begin{document}

\title{Counting And Confusion: Bayesian Rate Estimation With Multiple
  Populations}

\author{Will M. Farr} \email{wfarr@star.sr.bham.ac.uk}
\homepage{http://faculty.wcas.northwestern.edu/will-farr/}
\affiliation{Center for Interdisciplinary Exploration and Research in
  Astrophysics\\ Department of Physics and Astronomy\\ Northwestern
  University, 2145 Sheridan Road, Evanston, IL 60208}
\affiliation{School of
  Physics and Astronomy\\University of Birmingham\\Edgbaston B15 2TT
  Birmingham\\United Kingdom}

\author{Jonathan R. Gair} \email{jrg23@cam.ac.uk}
\affiliation{Institute of Astronomy\\University of
  Cambridge\\Madingley Road, Cambridge CB3 0HA\\United Kingdom}

\author{Ilya Mandel} \email{imandel@star.sr.bham.ac.uk}
\homepage{http://www.sr.bham.ac.uk/~imandel} \affiliation{School of
  Physics and Astronomy\\University of Birmingham\\Edgbaston B15 2TT
  Birmingham\\United Kingdom}
  
\author{Curt Cutler} \email{Curt.J.Cutler@jpl.nasa.gov}
\affiliation{Jet Propulsion Lab \\ 4800 Oak Grove Dr \\ Pasadena, CA
  91109} \affiliation{Theoretical Astrophysics \\ California Institute
  of Technology \\ Pasadena, CA 91125}

\begin{abstract}
  We show how to obtain a Bayesian estimate of the rates or numbers of
  signal and background events from a set of events when the shapes of
  the signal and background distributions are known, can be estimated,
  or approximated; our method works well even if the foreground and
  background event distributions overlap significantly and the nature
  of any individual event cannot be determined with any certainty.  We
  give examples of determining the rates of gravitational-wave events
  in the presence of background triggers from a template bank when
  noise parameters are known and/or can be fit from the trigger data.
  We also give an example of determining globular-cluster shape,
  location, and density from an observation of a stellar field that
  contains a non-uniform background density of stars superimposed on
  the cluster stars.
\end{abstract}

\maketitle

\section{Introduction}

The task of estimating rates of events when a mixture of foreground and background
events is present in data is a common one in physical and
astrophysical applications.  This problem comes up, among others, in
gravitational-wave data analysis
\cite[e.g.,][]{s5-lowmass,s5-highmass,s5-highmass-erratum,s5-lowmass-2,s6-lowmass,s6-highmass}
and in astronomical observations of a field of objects of mixed
provenance \cite{DeGennaro2009,Koposov2010}.  In this paper, we introduce a robust
formalism for estimating event rates from the data when the shape of
foreground and background distributions are known (or parameterized),
but the provenance of individual events as either background or
foreground is unknown.  

We use a Bayesian approach and consider all available data to ensure
that the inferred rates are both unbiased and maximally constrained in
the presence of limited observations.  Bayes' theorem yields the
\emph{posterior} probability density function on a set of parameters,
$\vec{\theta}$, given the observed data, $d$, under a model $M$:
\begin{equation}
p(\vec{\theta} | d, M) = \frac{p(\vec{\theta}|M)
  p(d|\vec{\theta},M)}{p(d|M)},
\end{equation}
where $p(\vec{\theta}|M)$ are the \emph{prior} probabilities of the
model parameters, $p(d|\vec{\theta},M)$ is the \emph{likelihood} of
obtaining the data given a particular choice of parameters, and the
normalizing factor $p(d|M)$ is known as the \emph{evidence}.

Two alternative approaches to rate estimation have been suggested and
are commonly used.  One, known as the \emph{loudest-event statistic}
\cite{Biswas2009,Brady2004,Brady2008}, uses only the information from
the highest-ranked event in the data to infer the rate distribution.
This approach has been used successfully
\cite{s5-lowmass,s5-lowmass-2,s5-highmass,s5-highmass-erratum,s6-lowmass,s6-highmass}
when the number of loud foreground events is small (typically zero or
one) to obtain upper limits on foreground rates.  However, the
loudest-event statistic ignores all events except the loudest one, and
so suffers from an unnecessary loss of information; therefore, we
expect it to yield a much larger variance than strictly necessary when
multiple events are present in the data.  In practice, the
loudest-event statistic is typically applied repeatedly to multiple
``chunks'' of data, using the estimated rate posterior from each chunk
as a rate prior for the next chunk's analysis
\cite{s5-lowmass-2,s5-highmass,s5-highmass-erratum}.  Even when used
in this mode, the method discards information, with the amount of
information loss depending on the (arbitrary) division of the data
into chunks.

Another possible approach is based on the use of only loud,
``gold-plated'' events, ones which are certain (or nearly certain) to
come from the foreground, to derive rates.  We refer to this approach
as the \emph{foreground-dominated statistic}.  The
foreground-dominated statistic may yield accurate results when the
foreground and background are cleanly separated, at least for the
loudest events, and the number of such loud events is sufficiently
large.  However, it cannot properly account for marginal events.  In
addition, the results of the method are very sensitive to
contamination by the background events, and therefore the method
requires a careful choice of \emph{threshold} or reliable
\emph{membership} information to distinguish foregrounds and
backgrounds for individual events.  While either the loudest-event
statistic or the foreground-dominated statistic can approach the
accuracy of our proposed method in specific regimes, both are
suboptimal in a general case.

Ref.~\cite{Messenger2012} considered the problem of determining an
intrinsic rate and population parameters in the presence of missing
data, either due to thresholding, poor sensitivity, or contamination
from noise events.  The approach is complementary to ours: we consider
the problem of accurately counting the events of different classes
present in a dataset, while Ref.~\cite{Messenger2012} deals with
translating such counts into physical rates by properly accounting for
the selection effects on the data set.  

Our key results appear in Eq.~(\ref{eq:posterior}), which provides the
joint posterior probability distribution on the foreground and
background rates and shape parameters and the provenance of individual
events as either foreground or background.
Eq.~(\ref{eq:rate-shape-posterior}) is a marginalized version of
Eq.~(\ref{eq:posterior}), useful when the provenance of individual
events is not relevant.  In practice, these posteriors are best
sampled with stochastic techniques such as Markov chain Monte Carlo.

In order to demonstrate our method, we consider three different
examples.  The first two come from the field of gravitational-wave
data analysis, but could equally arise in any application that employs
matched filtering \cite{findchirppaper} to extract weak signals with
known shapes from the data.  The last example considers the case of a
globular cluster on a background of field stars.  Throughout, we
compare the results obtained with our technique to the loudest-event
and foreground-dominated statistics, which make use of a limited
subset of the available information.

\section{Model}\label{sec:model}

We first consider one-dimensional data, but will generalize to the
multidimensional case below.  We assume that we are presented with a
data set of $N$ events that exceed a pre-specified threshold in
ranking statistic, $x_{\rm min}$.  Each event may be due to either a
signal of interest or an uninteresting background.  Each event is
associated with a ranking statistic, $x$.  Our data set therefore
consists of the ranking statistics for the set of events:
\begin{equation}
  d = \mathset{ x_i | i = 1, \ldots, N } .
\end{equation}
The number of events $N$ is also part of the observed data, but we
separate out $N$ and the observed ranking statistics, $d$, for
convenience. We can choose how to label our events. Ultimately we will
label the events in order of ranking statistic, i.e., $x_1 < x_2 <
\cdots < x_N$, but some of the derivations that follow are simpler if
the events are ordered by time of arrival (i.e.\ randomly with respect
to the $x_i$). We will use $d$ to denote ranking statistic-ordered
events, and $d_{\rm to}$ to denote time-ordered events.

We assume that both the foreground and background events are samples
from an inhomogeneous Poisson process with respective differential
rates
\begin{equation}
  \frac{dN_f}{dx} = f(x, \theta)
\end{equation}
and
\begin{equation}
  \frac{dN_b}{dx} = b(x, \theta),
\end{equation}
where the $\theta$ argument represents additional ``shape'' parameters
that may affect the distribution, and for which we will eventually
fit.  The cumulative rates of the two processes are therefore
\begin{equation}
  F(x,\theta) \equiv \int_{-\infty}^x ds\, f(s, \theta)
\end{equation}
and
\begin{equation}
  B(x,\theta) \equiv \int_{-\infty}^x ds\, b(s,\theta).
\end{equation}
The assumption that the foreground and background events form an
inhomogeneous Poisson process implies
\begin{enumerate}
\item \label{prop:Poisson} The number of events in any range of
  ranking statistics, $x \in [x_1, x_2]$ is Poisson distributed with
  rate $F(x_2, \theta) - F(x_1,\theta)$ or $B(x_2,\theta) -
  B(x_1,\theta)$.
\item The numbers of events in non-overlapping ranges of ranking
  statistics are independent.
\item The probability of exactly one foreground event between $x$ and
  $x+h$ is given by
  \begin{equation}
    P(n = 1 \in [x, x+h]) = f(x,\theta) h + \order{h^2}.
  \end{equation}
  and similarly for background events.
\item The probability of two or more events in a small range of
  ranking statistic is negligible
  \begin{equation}
    P(n = 2 \in [x, x+h]) = \order{h^2}.
  \end{equation}
\end{enumerate}
The foreground and background rates can in general depend on several
parameters; the goal of our analysis is to determine the posterior
probability distributions for these parameters that are implied by the
data.  At the least, we will want to know the overall dimensionless
amplitude of the foreground and background rates.  Let
\begin{equation}
  f(x,\theta) = R_f \hat{f}(x,\theta'),
\end{equation}
and
\begin{equation}
  b(x, \theta) = R_b \hat{b}(x, \theta'),
\end{equation}
where $\hat{F}(\infty, \theta') = \hat{B}(\infty, \theta') = 1$, and
$\theta' = \theta \setminus \{R_{f}, R_{b} \}$.  Then $R_f \equiv
F(\infty,\theta)$ and $R_b \equiv B(\infty,\theta)$ are the total
number of foreground and background events expected and $\hat{f}(x,
\theta')$ and $\hat{b}(x, \theta')$ are the likelihood of obtaining an
event with ranking statistic $x$ under the foreground and background
distributions.  In what follows, we will drop the prime, using
$\theta$ to denote all parameters of the rate distributions except
$R_f$ and $R_b$.

We do not know a priori which of the events are foreground and which
are background.  For each event, we introduce a flag, $g_i$, which is
either 0 (background) or 1 (foreground).  These ``state'' flags are
parameters in our model, along with $R_f$, $R_b$, and $\theta$.  We
can marginalize over our uncertainty in the state of any given event
by summing posteriors over $g_i = \mathset{0,1}$.

Assuming time-ordered data, $d_{\rm to}$, in the following, Bayes'
theorem relates the posterior probability of the state flags, rates,
and shape parameters, $p\left(\mathset{g_i}, R_f, R_b, \theta | d_{\rm
  to}, N \right)$, the likelihood of the data, $p\left( d_{\rm to} |
\mathset{g_i}, N, R_f, R_b, \theta\right)$, and the prior probability
of state flags, rates and shape parameters before any data are
obtained, $p\left( \mathset{g_i}, N, R_f, R_b, \theta\right)$:
\begin{multline}
  \label{eq:Bayes}
  p\left(\mathset{g_i}, R_f, R_b, \theta | d_{\rm to}, N \right) \\ =
  \frac{p\left( d_{\rm to} | \mathset{g_i}, N, R_f, R_b, \theta\right)
    p\left(\mathset{g_i}, N, R_f, R_b, \theta\right)}{p(d_{\rm to},
    N)}.
\end{multline}
The normalization constant, called the evidence, $p(d_{\rm to}, N)$,
is independent of the state flags, rates, and shape parameters.

Each foreground event is drawn from the probability distribution
$\hat{f}$ and each background event is drawn from the probability
distribution $\hat{b}$.  The events are independent of each other.
Therefore, the likelihood of the data is
\begin{multline}
  p\left( d_{\rm to} | \mathset{g_i}, N, R_f, R_b, \theta\right) \\ =
  \left[ \prod_{\mathset{i | g_i = 1}} \hat{f}\left(x_i, \theta\right)
    \right] \left[ \prod_{\mathset{i | g_i = 0}} \hat{b}\left( x_i,
    \theta\right) \right].
\end{multline}
This is the probability that the first observed event is a
fore/background event (if $g_1=1,0$) with ranking statistic $x_1$ and
the second observed event is a fore/background event (if $g_2=1,0$)
with ranking statistic $x_2$, etc. If the events are ordered by ranking
statistic the corresponding expression is more complicated, since
$x_1$ is now the event from foreground or background with the smallest
ranking statistic, etc. We will return to the statistic-ordered case
later.

The prior distribution can be factorized as
\begin{multline}
  \label{eq:combined-flag-rate-prior}
  p\left(\mathset{g_i}, N, R_f, R_b, \theta\right) \\ = p\left(
  \mathset{g_i} | N, R_f, R_b\right) p\left(N |R_f, R_b\right)
  p\left(R_f, R_b, \theta\right) \\ = p\left( \mathset{g_i},N | R_f,
  R_b\right) p\left(R_f, R_b, \theta\right).
 \end{multline}
 
The probability that the $i$'th state flag is $g_i=1$ is given by
$R_f/(R_f+R_b)$, while the probability that it is zero is
$R_b/(R_f+R_b)$, provided the data are time-ordered as we have
assumed.  Then
\begin{multline}
p\left( \mathset{g_i} | N, R_f, R_b\right) \\ = \prod_{\mathset{i|g_i=1}}
\left(\frac{R_f}{R_f+R_b}\right) \prod_{\mathset{i|g_i=0}}
\left(\frac{R_b}{R_f+R_b}\right) \\ =
\left(\frac{R_f}{R_f+R_b}\right)^{N_f}
\left(\frac{R_b}{R_f+R_b}\right)^{N_b},
\end{multline}
where $N_f$ and $N_b$ are the numbers of foreground and background
flags, $N_f+N_b=N$.  Meanwhile,
\begin{equation}
p\left(N |R_f, R_b\right) = \frac{\left(R_f+R_b\right)^N}{N!}
e^{-(R_f+R_b)},
\end{equation}
since the distribution of total event number is a Poisson process with
rate $R_f+R_b$.  Combining these yields the conditional probability of
the flags on the rates:
\begin{equation}
  \label{eq:flag-conditional-prior}
  p\left(\mathset{g_i},N | R_f, R_b\right) = \frac{R_f^{N_f}
    R_b^{N_b}}{N!} \exp\left[ - \left(R_f + R_b\right) \right].
\end{equation}

The last term in Eq.~\eqref{eq:combined-flag-rate-prior} is a
traditional prior.  Because the rate parameters enter the posterior in
the same form as Poisson rates, we choose here the Poisson Jeffreys
prior on rates \citep{Jeffreys1946}, independent of the shape
parameters
\begin{equation}
  p\left( R_f, R_b, \theta\right) = \alpha \frac{1}{\sqrt{R_f R_b}}
  p(\theta),
\end{equation}
where $\alpha$ is a normalization constant; but of course other
choices are possible.  This choice has the advantage that the prior is
normalizable as $R_f, R_b \to 0$, and the exponentials in
Eq.~\eqref{eq:flag-conditional-prior} regularize the posterior as
$R_f, R_b \to \infty$.

Putting everything together, the posterior is
\begin{multline}
  \label{eq:posterior}
  p\left( \mathset{g_i}, R_f, R_b, \theta | d_{\rm to},N \right) =
  \frac{\alpha}{p(d_{\rm to},N)\,N!} \\ \times \left[ \prod_{\mathset{i | g_i =
        1}} R_f \hat{f}\left(x_i, \theta\right) \right] \left[
    \prod_{\mathset{i | g_i = 0}} R_b \hat{b}\left( x_i, \theta\right)
    \right] \\ \times \exp\left[ - \left(R_f + R_b\right) \right]
  \frac{p(\theta)}{\sqrt{R_f R_b}}.
\end{multline}
When sampling the posterior, the first term, which is independent of
the parameters of interest, can be omitted and the equals sign
replaced by proportionality; however, we have kept this term
explicitly so that we can see the equivalence to ranking-statistic
ordered data. Once data have been observed, there is a unique loudness
ordering and time ordering of those events, and so there is a one to
one correspondence between a time-ordered posterior $p\left(
\mathset{g_i}, R_f, R_b, \theta | d_{\rm to},N \right)$ and the
corresponding statistic-ordered posterior $ p\left( \mathset{g_i},
R_f, R_b, \theta | d,N \right)$, which means $p\left( \mathset{g_i},
R_f, R_b, \theta | d,N \right) = p\left( \mathset{g_i}, R_f, R_b,
\theta | d_{\rm to},N \right)$. However, the evidence $p(d,N) =
N!\,p(d_{\rm to},N)$, since there are $N!$ ways in which $N$ events
with a given set of ranking statistics can be ordered in time.

The ranking-statistic ordered posterior can be computed directly by
assuming that the flags, $\mathset{g_i}$, are un-observed data and
treating the sets $\mathset{x_i | g_i = 1}$ and $\mathset{x_i | g_i =
  0}$ as samples from an inhomogeneous Poisson process.  For an
inhomogeneous Poisson process with rate function $r(y)$ (cumulative
rate $R(y)$), the likelihood of a set of samples $\mathset{y_i}$ is
given by
\begin{multline}
  p\left( \mathset{y_i} | r \right) \,{\rm d}^Ny_i = P\left(
  \textnormal{zero events below } y_1 \right) \\ \times P\left(
  \textnormal{one event between } y_1 \textnormal{ and } y_1 + {\rm
    d}y_1 \right) \\ \times P\left( \textnormal{zero events between }
  y_1+{\rm d}y_1 \textnormal{ and } y_2 \right) \ldots,
\end{multline}
so 
\begin{multline}
p\left(\mathset{y_i} | r \right) = \lim_{\delta y_i \to 0} \exp\left[ -
    R\left(y_1\right) \right] \left[ r\left( y_1 \right) +
    \order{\delta y_1} \right] \\ \times \exp\left[ - \left[ R\left( y_2
      \right) - R\left( y_1 + \delta y_1\right) \right] \right] \times
  \ldots \\ = \left[\prod_i r\left( y_i \right)\right] \exp\left[ -
    R\left( \infty \right) \right].
\end{multline}
Applying this once to the foreground samples, once to the background
samples and taking the product, we obtain $p(d,\mathset{g_i}, N | R_f,
R_b, \theta)$ and thus $p(\mathset{g_i}, R_f, R_b, \theta | d,N) =
p(d,\mathset{g_i}, N | R_f, R_b, \theta)\ p(R_f, R_b, \theta)\, /\,
p(d,N)$. With the identification $p(d,N) = N!\,p(d_{\rm to},N)$, as
justified above, we reproduce Eq.~\eqref{eq:posterior}.

We can marginalize the posterior over the flags, $g_i$, obtaining
\begin{multline}
  \label{eq:rate-shape-posterior}
  p\left( R_f, R_b, \theta | d, N \right) = \sum_{\mathset{g_i} \in
    \mathset{0,1}^N} p\left( \mathset{g_i}, R_f, R_b, \theta | d, N
  \right) \\ \propto \prod_{i} \left[ R_f \hat{f}\left(x_i, \theta\right)
    + R_b \hat{b}\left( x_i, \theta\right) \right] \\ \times \exp\left[-\left(
    R_f + R_b \right) \right] \frac{p(\theta)}{\sqrt{R_f R_b}}.
\end{multline}
This expression is useful if we are only interested in rates and not
the probability that any particular event is foreground or background.
Unlike the full posterior (Eq.~\eqref{eq:posterior}),
Eq.~\eqref{eq:rate-shape-posterior} contains only continuous
parameters. We note that the terms that depend on the overall rate
parameters, $R_b$ or $R_f$, are of the form $R_b^{n-1/2} \exp(-R_b)$
and so marginalization over either $R_b$ or $R_f$ can be achieved
analytically using
\begin{equation}
I_n = \int_0^{\infty} x^{n-\frac{1}{2}} \, {\rm e}^{-x} {\rm d} x =
\frac{(2n-1)!!}{2^n} \sqrt{\pi}
\end{equation}
using the usual notation $(2n-1)!! \equiv (2n-1)(2n-3)\cdots1$.

Eq.~\eqref{eq:posterior} is unchanged if the ranking statistic is
multi-dimensional; in this case, the rates are
\begin{equation}
  R_f = \int d^k \vec{x} \, f(x, \theta)
\end{equation}
and
\begin{equation}
  R_b = \int d^k \vec{x} \, b(x, \theta),
\end{equation}
where $f$ and $b$ are rate densities on the $k$-dimensional space of
ranking statistics.  We give an example of fitting for
multi-dimensional rate densities in \S~\ref{sec:star-cluster}.

\section{Comparison to Other Rate Estimation Methods}

It is informative to relate these results to two other methods for
estimating the foreground rate parameter --- the loudest event
statistic and the foreground-dominated statistic.

\subsection{Loudest event statistic}
If we were to include only the $k$ loudest events in the posterior
distribution, rather than all observed events, the posterior
(Eq.~\eqref{eq:posterior}) would be modified by an additional factor
of $\exp[R_f \hat{F}(x_{N-k+1},\theta) + R_b
  \hat{B}(x_{N-k+1},\theta)]$, where we have assumed events are
ordered by loudness, so that $x_{N-k+1}$ is the $k$-th loudest
event. This term accounts for the data-dependent threshold that a
loudest event statistic employs.

For the usual $k=1$ case \citep{Biswas2009}, the marginalized
posterior (Eq.~\eqref{eq:rate-shape-posterior}) becomes
\begin{multline}\label{LEeq1}
p_{\rm LE}\left( R_f, R_b, \theta | d \right) \propto \left(R_f
\hat{f}(x_N,\theta) + R_b \hat{b}(x_N,\theta)\right) \\ \times \exp\left[-\left(
  R_f (1-\hat{F}(x_N,\theta)) + R_b (1-\hat{B}(x_N,\theta)) \right)
  \right] \\ \times \frac{p(\theta)}{\sqrt{R_f R_b}}.
\end{multline}
where $x_N$ denotes the loudness of the loudest event, and $R_f$ and
$R_b$ are the number of events expected \emph{above our original
  threshold} (so, for example, $R_f (1 - \hat{F}(x_N,\theta))$ is the
number of foreground events expected above loudness $x_N$).  In the
loudest event statistic paper~\citep{Biswas2009}, the authors assume
the background distribution and rate are known, which corresponds to
using a narrow prior on $R_b$. They further assume a flat prior (in
the absence of other experimental data) on $R_f$ and that the
foreground and background distributions do not depend on any unknown
free parameters. With these assumptions, the posterior on $R_f$,
Eq.~\eqref{LEeq1}, is modified to
\begin{multline}
p_{\rm LE}(R_f|d) \propto  \left(R_f
\hat{f}(x_N) + R_b \hat{b}(x_N)\right) \\ \times \exp\left[-\left(
  R_f (1-\hat{F}(x_N)) + R_b (1-\hat{B}(x_N)) \right)
  \right] .
\end{multline}
Integrating over $R_f$ gives
\begin{multline}
\int_0^\infty p_{\rm LE}(R_f|d) {\rm d}R_f =  \frac{R_b \hat{b}(x_N)}{(1-\hat{F}(x_N))} {\rm e}^{-(1-\hat{B}(x_N)) R_b}\\
\times \left(\frac{\hat{f}(x_N)}{(1-\hat{F}(x_N)) R_b \hat{b}(x_N)} + 1 \right)
\end{multline}
and so the normalised posterior is
\begin{multline}
p_{\rm LE}(R_f|d) = \frac{(1-\hat{F}(x_N))}{1+\Lambda} \left( 1+ R_f (1-\hat{F}(x_N))\Lambda \right) \\
\times \exp\left[-R_f (1-\hat{F}(x_N))\right]
\label{loudest-eq}
\end{multline}
in which we have defined
\begin{equation}
\Lambda \equiv \frac{\hat{f}(x_N)}{(1-\hat{F}(x_N)) R_b \hat{b}(x_N)}.
\end{equation}
With the further identification $\mu \equiv R_f$ and $\hat{\epsilon} \equiv 1 - \hat{F}(x_N)$, this is Eq.~(14) of~\citep{Biswas2009} and we have shown how their parameter $\Lambda$ is related to the foreground and background distributions used here.

Returning now to Eq.~\eqref{LEeq1} and marginalizing over $R_b$, we obtain
\begin{multline}
p_{\rm LE}\left( R_f, \theta | d \right) \propto \left(
\frac{\hat{b}(x_N,\theta)}{2(1-\hat{B}(x_N,\theta))} + R_f
\hat{f}(x_N,\theta)\right) \\ \times \frac{\sqrt{\pi}\,
  p(\theta)}{\sqrt{1-\hat{B}(x_N,\theta)}\,\sqrt{R_f}}
\exp\left(-R_f(1-\hat{F}(x_N,\theta))\right) .
\end{multline}
This posterior has a maximum in $R_f$ at
\begin{multline}
R_f =
\frac{\hat{f}(x_N,\theta)-(1-\hat{F}(x_N,\theta))\tilde{b}(x_N,\theta)
  + \sqrt{g(x_N,\theta)}}{4\hat{f}(x_N,\theta)(1-\hat{F}(x_N,\theta))} \\
\mbox{where }  g(x_N,\theta) =\left(\hat{f}(x_N,\theta)-(1-\hat{F}(x_N,\theta))\tilde{b}(x_N,\theta)\right)^2
    \\ - 4\tilde{b}(x_N,\theta)(1-\hat{F}(x_N,\theta))\hat{f}(x_N,\theta)
\end{multline}
and $\tilde{b}(x_N,\theta) = \hat{b}(x_N,\theta) /(1-\hat{B}(x_N,\theta))$ and similarly for $\tilde{f}(x_N,\theta)$.

If $\tilde{b}(x_N,\theta) \ll \tilde{f}(x_N,\theta)$, we obtain the
result $(1- \hat{F}(x_N,\theta))\,R_f \approx 1/2$. This can be
understood as the statement that the rate of foreground events with
ranking statistic greater than $x_N$, $(1- \hat{F}(x_N,\theta))\,R_f$,
is of order $1$, as expected. However, $\tilde{b}(x_N,\theta) = -{\rm
  d} [\ln(1-\hat{B}(x,\theta))]/{\rm d}x$ and $(1-\hat{B}(x,\theta))
\rightarrow 0$ as $x \rightarrow \infty$, so this term may be
divergent and for many reasonable examples, we will find
$\tilde{b}(x_N,\theta) \gg \tilde{f}(x_N,\theta)$, in which case the
posterior on $R_f$ is peaked at $0$. This issue highlights the problem
with using a loudest-event statistic with an improper prior on the
background rate $R_b$. No matter how improbable an event with $x=x_N$
is under the background distribution, it can become likely that the
event at $x_N$ is from the background distribution by taking the
background rate to be sufficiently large. Although this predicts many
more events with $x < x_N$, by using only the loudest event we do not
incorporate the information that no such events are seen. This problem
is avoided in the new framework described here, since we use all
events detected above threshold and combined rates, $R_f+R_b$,
significantly greater than the total number of observed events are
strongly disfavored.  

This problem can also be avoided in the context of the loudest-event
framework by even very weak prior information on the background rate,
$R_b$, of the kind present in nearly all experiments.  For example, we
can include an upper limit on the rate, $R_{\rm max}$, in the prior
for $R_b$.
\begin{widetext}
The marginalized distribution for the foreground rate then becomes
\begin{eqnarray}
p_{\rm LE}\left( R_f, \theta | d \right) &\propto& \left(
\frac{\hat{b}(x_N,\theta)}{(1-\hat{B}(x_N,\theta))^{\frac{3}{2}}}
\left[\frac{\sqrt{\pi}}{2}{\rm
    erf}\left(\sqrt{(1-\hat{B}(x_N,\theta))\,R_{\rm max}}\right) -
  \sqrt{(1-\hat{B}(x_N,\theta))\,R_{\rm max}} \, {\rm
    e}^{-(1-\hat{B}(x_N,\theta)) \, R_{\rm max}}\right]
\right. \nonumber \\ && \left.+ R_f
\hat{f}(x_N,\theta)\frac{\sqrt{\pi}\,{\rm
    erf}\left(\sqrt{(1-\hat{B}(x_N,\theta))\,R_{\rm
      max}}\right)}{\sqrt{(1-\hat{B}(x_N,\theta))}}\right)\frac{p(\theta)}{\sqrt{R_f}}
\exp\left(-R_f(1-\hat{F}(x_N,\theta))\right) \label{LEwithRmax},
\end{eqnarray}
where ${\rm erf}(x)$ is the error function, defined in the usual way
${\rm erf}(x) = (2/\sqrt{\pi}) \int_0^x \exp(-u^2){\rm d}u$. If
$(1-\hat{B}(x_N,\theta)) \, R_{\rm max} \ll 1$, Eq.~(\ref{LEwithRmax})
can be approximated by
\begin{equation}
p_{\rm LE}\left( R_f, \theta | d \right) \propto \left( \frac{R_{\rm
    max}}{3} \hat{b}(x_N,\theta) + R_f
\hat{f}(x_N,\theta)\right)\frac{p(\theta)}{\sqrt{R_f}}
\exp\left(-R_f(1-\hat{F}(x_N,\theta))\right)
\end{equation}
and if $\hat{f}(x_N,\theta) \gg R_{\rm max} \hat{b}(x_N,\theta)$ we
find the same result as before, $(1- \hat{F}(x_N,\theta))\,R_f \approx
1/2$.
\end{widetext}

\subsection{Foreground dominated statistic}
If we set the threshold for including an event, $x_{\rm min}$,
sufficiently high, we can ensure that $\hat{f}(x_i,\theta) \gg
\hat{b}(x_i,\theta)$ for all ranking statistics $x_i$ in the data
set.  If we can further be confident that $R_f \hat{f}(x_i,\theta) \gg R_b \hat{b}(x_i,\theta)$ for all events, then the posterior can be approximated by
\begin{multline}
p_{\rm FD}\left( R_f, R_b, \theta | d \right) \\ \propto \prod_{i} \left[
  \hat{f}\left(x_i, \theta\right)\right] R_f^N \exp\left[-\left( R_f +
  R_b \right) \right] \frac{p(\theta)}{\sqrt{R_f R_b}}.
\end{multline}
Note that these are posteriors on the number of events expected \emph{above the
threshold $x_{\rm min}$}.  The threshold choice for the foreground-dominated statistic could be different from the threshold choice applied elsewhere.  
If the rates are estimated accurately, then a rate estimate $R_{f,1}$
above threshold $x_{\rm min}=x_1$ can be converted into a rate
estimate $R_{f,2}$ above threshold $x_{\rm min}=x_2$ via $R_{f,1}
(1-\hat{F}(x_2,\theta)) = R_{f,2} (1-\hat{F}(x_1,\theta))$; however,
rate point estimates based on thresholding can have significant
fluctuations, as discussed in the following section.

Normalization over $R_b$ gives a constant factor and the posterior on
the foreground rate becomes
\begin{multline}
p_{\rm FD}\left( R_f, \theta | d \right) \\ \propto \prod_{i} \left[
  \hat{f}\left(x_i, \theta\right)\right] R_f^{N-\frac{1}{2}}
\exp\left[-R_f \right] p(\theta).
\end{multline}
Ignoring the dependence on $\theta$, this is peaked at a rate $R_f =
N-1/2$, so we have the expected result that, in the foreground
dominated regime, the rate is approximately equal to the number of
events observed (the $1/2$ comes from our use of the Jeffreys prior on
the rate).  

\section{Thresholding}
This paper is concerned with Bayesian rate estimates based on lists of
events.  Ideally, the lists should contain all events in the data set.
However, for experimental or computational reasons one may wish to
restrict the events to only those above some loudness threshold; in
some cases the rate of foreground or background events, or both, is
even expected to diverge at certain loudnesses.  In this subsection we
address the question of how the rate estimate depends on the threshold
value.  For a discussion of selection effects, of which thresholding
is but one, on the estimate of physical rates, see
Ref.~\cite{Messenger2012}.

To begin with, we recall the well-known fact that the Bayesian
estimator is unbiased, in the following sense.  For simplicity, assume
that the model consists of a single rate parameter $R$, with prior
distribution $p(R)$.  Consider an ensemble of data sets whose
distribution is consistent with that prior; i.e., such that $p(d)$ is
given by
\begin{equation}
 p(d) = \int{ p(d|R)\,p(R)\, dR}\, .  
\end{equation}
For each data set in the ensemble, compute the Bayesian estimator 
for the mean of the posterior $R_{B} = \int{ R p(R|d) dR}$.  Then it is immediate that
\begin{equation}
\int{R_{B}(d)\, p(d)\,d(d) }= \int{ R\, p(R) \,dR},
\end{equation}
i.e.\ the data-weighted average of the Bayesian estimator $R_{B}$
equals the prior-weighted average $R$.  Therefore all threshold values
will yield, on average, the same point estimate of the rate.  However
this equality of averages does {\it not} imply that all threshold
values yield the same information. In general, as the threshold is
lowered to include more events, the error bar on the estimate shrinks.
In this subsection we give quantitative illustrations of how the error
bar shrinks when the threshold is lowered.

Consider the following model problem.  Let $p(x) = b(x) + f(x) = R_b
\hat{b}(x) + R_f \hat{f}(x)$ be the rate density of events (of both
foreground and background type) per unit loudness.  Here we will
assume that the background is normally-distributed in loudness, so
that $b$ has the form
\begin{equation}
b(x) = \Gamma_b \exp \left(- \frac{x^2}{2} \right).
\end{equation}
We find it useful to define $x_1$ as the loudness such that a data set
will have on average a single noise event louder than $x_1$; i.e.,
such that 
\begin{equation}
 \int_{x_1}^{\infty} b(x) dx = R_b - B\left(x_1\right) = 1.
\end{equation}
This condition fixes 
\begin{equation}
  \Gamma_b = \left[ \sqrt{\frac{\pi}{2}} \erfc\left(
    \frac{x_1}{\sqrt{2}} \right) \right]^{-1},
\end{equation}
while $R_b$ will depend on the threshold, $x_\mathrm{th}$, as 
\begin{equation}
  R_b = \frac{\erfc\left( \frac{x_\mathrm{th}}{\sqrt{2}}
    \right)}{\erfc\left( \frac{x_1}{\sqrt{2}} \right)}.
\end{equation}

Let the foreground distribution follow a power law in loudness (this
is, for example, the distribution of SNR for gravitational wave events
from uniformly-distributed sources in a single detector)
\begin{equation}
\label{eq:ps}
f(x) = 3 \Gamma_f \frac{x_1^3}{x^4},
\end{equation}
where $\Gamma_f = R_f - F\left(x_1\right)$ is the mean number of
foreground events with $x > x_1$.  The overall foreground rate is
given by 
\begin{equation}
  \label{eq:foreground-from-lambda}
  R_f = \Gamma_f \frac{x_1^3}{x_\mathrm{th}^3}.
\end{equation}

We can write the full $p(x)$ as
\begin{equation}
\label{eq:ptot} 
p(x) = \left[\sqrt{\frac{\pi}{2}}
  \erfc\left(\frac{x_1}{\sqrt{2}}\right)\right]^{-1}\exp\left[-\frac{x^2}{2}\right]
+ 3 \Gamma_f \frac{x_1^3}{x^4}
\end{equation}

For any pair $(x_1, \Gamma_f)$, it is straightforward to construct
random event lists drawn from the corresponding $p(x)$, and
straightforward to apply a threshold by ``throwing away'' all events
with $x$ less than the threshold value $x_\mathrm{th}$.  If $\Gamma_f
\gg 1$, then we are in the foreground-dominated regime at $x = x_1$,
if $\Gamma_f \ll 1$ we are in the background-dominated regime, and if
$\Gamma_f \sim 1$ the foreground and background counts above $x_1$
are about equal.  For any thresholded event list, we use
Eq.~(\ref{eq:rate-shape-posterior}) to construct the probability
density $p(R_f|d)$.  For that event list, we define the foreground
rate uncertainty, $\Delta R_f$, by
\begin{equation}
\label{eq:delta-R-definition}
(\Delta R_f)^2 \equiv \int{(R_f - R_f^\mathrm{true})^2 \, p(R_f|d)\, dR_f },
\end{equation}
where $R_f^\mathrm{true}$ is given by
Eq.~\eqref{eq:foreground-from-lambda}.

\begin{figure}
  \includegraphics[width=\columnwidth]{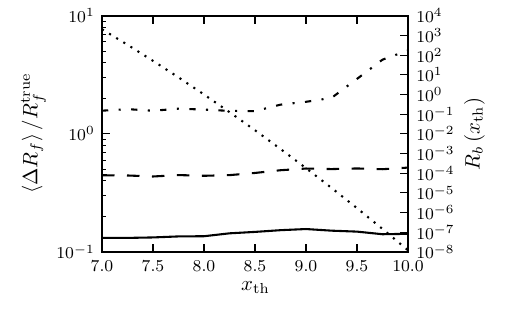}
  \caption{\label{fig:threshold-rate-dependence} The mean foreground
    rate uncertainty, Eq.~\eqref{eq:delta-R-definition}, as a function
    of threshold for data sets with $\Gamma_f = 100$ (solid line),
    $\Gamma_f = 10$ (dashed line), and $\Gamma_f = 1$ (dash-dotted
    line).  Recall that $\Gamma_f$ is the mean number of foreground
    events above $x_\mathrm{th}=x_1= 8$.  The total background rate,
    $R_b\left( x_\mathrm{th} \right)$, is shown by the dotted line; we
    fix $R_b \left( x_\mathrm{th} = x_1 = 8\right) = 1$, so on average
    there is one background event above $x = 8$.  For $x_\mathrm{th}
    \gtrsim x_1$, increasing the threshold tends to increase the
    foreground rate uncertainty because the rate is
    foreground-dominated and fewer events are included in the data
    set.  For $x_\mathrm{th} \lesssim x_1$, the background rate
    dominates at small loudness, and the foreground rate uncertainty
    asymptotes to the counting error on the events that stand out from
    the background, $\Delta R_f / R_f \simeq 1/\sqrt{\Gamma_f}$.}
\end{figure}

Figure \ref{fig:threshold-rate-dependence} illustrates how the mean
fractional foreground uncertainty, $\left\langle \Delta R_f
\right\rangle / R_f$, varies with the threshold value $x_\mathrm{th}$
for the foreground-dominated and comparable-rate regime.  In all cases
we assumed that $x_1 = 8$.  For large thresholds, where $R_b \ll 1$,
increasing the threshold tends to increase the fractional uncertainty
on the foreground rate, since fewer foreground events are included in
the sample.  However, as the threshold passes into the
background-dominated regime, the uncertainty in the foreground rate
asymptotes to
\begin{equation}
  \frac{\Delta R_f}{R_f} \simeq \frac{1}{\sqrt{\Gamma_f}},
\end{equation}
which is the usual Poisson counting uncertainty on the events that
stand out from the background (those with $x \gtrsim x_1$).  Note that
this uncertainty applies even when the total number of background
events is orders of magnitude larger than the number of foreground
events.  When a threshold must be chosen, it is safest---in the sense
of producing the minimal foreground rate uncertainty---to choose the
threshold \emph{well into the background-dominated loudness regime};
the extra background events in the data set do not affect the estimate
of the foreground rate, and, when the background distribution is
parameterized, can help to better determine these parameters (see
\S~\ref{sec:gw-overlapping-template}).

Though we have only illustrated the behavior of the rate estimate
quantitatively for this specific example of foreground and background
rates, the conclusions hold in general.  Consider the Fisher
information matrix for the posterior distribution in
Eq.~\eqref{eq:rate-shape-posterior}.  For a model with parameters
$\mathset{\theta_i}$, the Fisher information matrix has components
\begin{equation}
  F_{ij} \equiv \left\langle \frac{\partial \log p\left( \theta | d
    \right)}{\partial \theta_i} \frac{\partial \log p\left( \theta | d
    \right)}{\partial \theta_j} \right\rangle,
\end{equation}
where the average is taken over the data distribution at fixed
$\theta$, $p(d | \theta)$.  The components of the Fisher information
matrix describe the maximum amount of information about the
corresponding parameters available in a given data set; the inverse of
the Fisher information matrix gives the Cramer-Rao bound on the
covariance matrix of unbiased estimators of $\theta$.  Though our
Bayesian analysis is not necessarily limited by the Cramer-Rao bound
(since estimators constructed from it need not be unbiased and can
also be affected by the prior), the Fisher information is indicative
of the influence of each measurement on the posterior.  For the
likelihood that enters Eq.~\eqref{eq:rate-shape-posterior}, the Fisher
information matrix is
\begin{multline}
  \mathbf{F} = \left( R_f + R_b \right) \\ \times \begin{pmatrix}
    \left \langle \left(\frac{\hat{f}}{R_f \hat{f} + R_b \hat{b}}\right)^2 \right
    \rangle & \left \langle \frac{\hat{f} \hat{b}}{\left( R_f \hat{f}
      + R_b \hat{b} \right)^2} \right\rangle \\
    \left \langle \frac{\hat{f} \hat{b}}{\left( R_f \hat{f}
      + R_b \hat{b} \right)^2} \right\rangle & \left \langle
    \left(\frac{\hat{b}}{R_f \hat{f} + R_b \hat{b}}\right)^2 \right 
    \rangle
  \end{pmatrix},
\end{multline}
where the expectation values are taken over the distributions
$\hat{f}$ and $\hat{b}$ (i.e.\ they are expectations for one event
from the combined rate distribution).  If the cross-terms are small,
then the Cramer-Rao bound on the uncertainty of $R_f$ will be given by
\begin{equation}
  \sigma_{R_f} \simeq \frac{1}{\sqrt{R_f + R_b}} \left[ \left\langle \left(
    \frac{\hat{f}}{R_f \hat{f} + R_b \hat{b}} \right)^2 \right\rangle\right]^{-1/2}
\end{equation}
Extending a threshold into regions where the factor 
\begin{equation}
  \left( \frac{\hat{f}}{R_f \hat{f} + R_b \hat{b}} \right)^2
\end{equation}
becomes small---that is, into background-dominated
regions---contributes little to reducing the overall uncertainty in the
foreground rate.  Thus, when the background distribution itself is of no interest and computational costs are high, the threshold does not need to be pushed into background-dominated regions in order to obtain an accurate foreground estimate. This is consistent with the behavior of the specific
example in Figure \ref{fig:threshold-rate-dependence}.  

\subsection{Extreme Sensitivity of the LE Rate Estimate to a Single, Unusually Loud Event}
Here we discuss a very unattractive feature of the Bayesian loudest
event estimate of $R$ \citep{Biswas2009}: a small percentage of the
time it will yield a very large over-estimate.

To explain this, we will use the same model as described in the
previous subsection, and we will begin with a very specific example.
Let $\Gamma_f = 1$, meaning that the expected number of actual events
with $x > x_1$ is one.  Then there is a $1/64$ chance ($1-\hat{F}(x_{LE}) \approx
1.6\%$) that the loudest event will have $x_{LE} > 4 x_1$.  Consider
this case, and let us also assume that there are no events (noise or
actual) with $x_1 < x < x_{LE}$.  

The loudest event estimate basically ``throws away'' the information that there are no events in this interval.  The maximum of the loudest-event-statistic posterior on $R_f$,  Eq.~(\ref{loudest-eq}), is at $R_f = \frac{\Lambda - 1}{ \Lambda (1-\hat{F}(x_{LE}))}$.  If the value of $\Lambda$ is sufficiently high at $x_1$ (and $\Lambda$ will be even greater at $x_{LE}$), then, for this data set, we would estimate $R_f \approx \frac{1}{1-\hat{F}(x_{LE})} \gtrsim 64$.  Thus, for our assumed shape of the foreground distribution, we will estimate the rate of events above $x_1$ to be $64$ times the true rate!  

Now, if the true rate really were $\Gamma_f = 64$, then the expected
number of events with $x > x_1$ would be $64$.  So in this case, the
loudest event estimate ignores the fact that there are $\sim 56 -72$
``missing'' events.  However a Bayesian estimate with $x_\mathrm{th}$
set to $x_1$ incorporates this information quite naturally, and so
(correctly) yields an estimated $\Gamma_f$ of order one.

\section{Examples}
\label{sec:GW-example}

In this section we present several examples of the application of our
framework to various rate estimation problems in the presence of
background.

\subsection{Gravitational Waves with Non-Overlapping Templates}
\label{sec:analytic-GW-example}

Suppose we attempt to detect gravitational wave signals in a data
stream by matched filtering in the frequency domain against a set of
$N$ template waveforms \citep[e.g.,][]{findchirppaper,s6-lowmass}.  We
use an extremely simplified model of such a search and the ensuing
analysis to demonstrate how our framework could be used in practice.

In our simplistic model, we suppose the data stream consists of
stationary Gaussian noise with a power spectral density $S(f)$
combined additively with some number of gravitational wave signals.
We assume that the signals are sufficiently rare that they do not
overlap in the data stream.  The signal-to-noise ratio (SNR) of a
template, $h(f)$, given data, $d(f)$, is
\begin{equation}
  \rho_h \equiv \frac{\left\langle h, d \right\rangle}{\sqrt{\left
      \langle h, h \right\rangle}},
\end{equation}
where $\left \langle \cdot \right\rangle$ denotes the noise-weighted
inner product:
\begin{equation}
  \left\langle a, b \right\rangle \equiv 4 \Re \int_0^\infty df\,
  \frac{a^*(f) b(f)}{S(f)}.
\end{equation}
We suppose for simplicity that the templates are sufficiently distinct
that
\begin{equation}
  \left\langle h_i, h_j \right\rangle \simeq \delta_{ij}.
\end{equation}
In the following subsection, we will generalize the model to
overlapping templates.  We rank candidate events by their maximum SNR
over the entire template bank,
\begin{equation}
  x \equiv \max_{h} \rho_h,
\end{equation}
and consider only events that have a maximum SNR above some threshold,
$x > \xmin$.

For a data stream of pure noise, $d(f) = n(f)$, the SNRs of the
templates are independent $N(0,1)$ random variables.  The background
ranking statistic (i.e.\ the maximum SNR over the template bank) then
has a cumulative distribution without thresholding of
\begin{equation}
  \hat{B}(x) = \left( \frac{1 + \erf\left( \frac{x}{\sqrt{2}}
    \right)}{2} \right)^N
\end{equation}
where ${\rm erf}(x)$ is the error function as before. Imposing the
threshold, $x > \xmin$, the cumulative distribution of the background
becomes
\begin{equation}
  \label{eq:analytic-background-rate}
  \hat{B}(x) = \frac{\left( 1 + \erf\left( \frac{x}{{\sqrt{2}}}
    \right) \right)^N - \left( 1 + \erf\left( \frac{\xmin}{{\sqrt{2}}}
    \right) \right)^N}{2^N - \left( 1 + \erf\left(
    \frac{\xmin}{{\sqrt{2}}} \right) \right)^N }
\end{equation}
for $x>\xmin$, $0$ otherwise.

The SNR of a gravitational-wave signal in an interferometric detector
scales as $1/d$ \citep{Finn1992}, where $d$ is the distance to the
source.  Ignoring cosmological effects, the number of sources scales
as $d^3$.  Thus, we expect that the foreground cumulative distribution
of events will follow
\begin{equation}
  \label{eq:analytic-foreground-rate}
  \hat{F}(x) = 1 - \frac{\xmin^3}{x^3}.
\end{equation}

Note that this scenario has no shape parameters $\theta$ for the
foreground and background distributions.

\begin{figure}
  \includegraphics[width=\columnwidth]{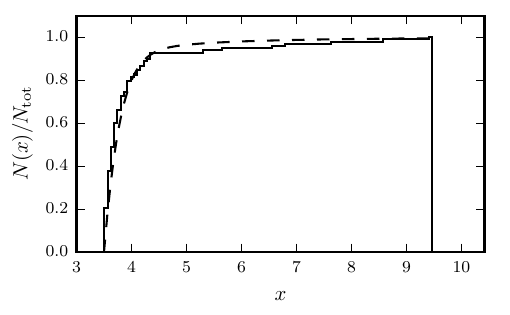}
  \caption{\label{fig:analytic-data-cumulative} The cumulative
    distribution of the ranking statistics for the synthetic data used
    to test the formalism on the model from \S
    \ref{sec:analytic-GW-example}.  The solid line gives the
    cumulative distribution of the synthetic data; the dashed line
    gives the theoretical cumulative distribution for the models in
    Eqs.~\eqref{eq:analytic-background-rate} and
    \eqref{eq:analytic-foreground-rate} combined with $R_f = 10.4$ and
    $R_b = 95.1$.}
\end{figure}

To demonstrate the effectiveness of our formalism, we applied it to a
synthetic data set with foreground and background distributions drawn
from Eqs.~\eqref{eq:analytic-background-rate} and
\eqref{eq:analytic-foreground-rate} using $x_\mathrm{min} = 3.5$, with
$R_f^\mathrm{true} = 10.4$ and $R_b^\mathrm{true} = 95.1$ and 1000
templates.  The synthetic data consisted of 13 foreground events and
85 background events; the cumulative distribution for the ranking
statistic of the synthetic data appears in Figure
\ref{fig:analytic-data-cumulative}.  We used a Markov chain Monte
Carlo simulation to draw samples of state flags and rates from the
joint posterior (Eq.~\eqref{eq:posterior}).

\begin{figure}
  \includegraphics[width=\columnwidth]{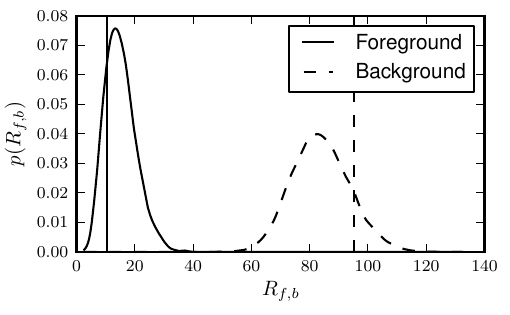}
  \caption{\label{fig:analytic-rate-recovery} The marginalized
    posterior densities for $R_f$ (solid line) and $R_b$ (dashed line)
    for the analytic model discussed in \S
    \ref{sec:analytic-GW-example}.  The vertical lines indicate the
    ``true'' values used to generate the synthetic data set.  Both the
    true foreground and background rates lie well within the
    probability envelope for $R_f$ and $R_b$.}
\end{figure}

\begin{figure}
  \includegraphics[width=\columnwidth]{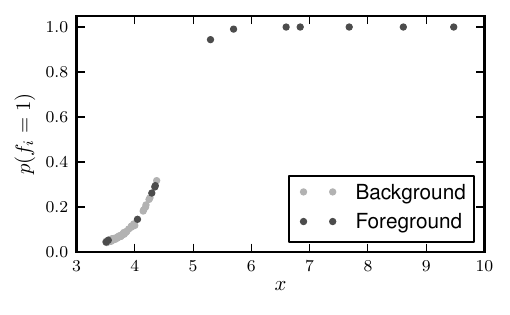}
  \caption{\label{fig:analytic-rate-foreground-probs} Foreground
    probability for each event in the synthetic data set of \S
    \ref{sec:analytic-GW-example} marginalized over all other
    parameters.  True foreground events are in dark grey, background
    events in light grey.  Even though our method cannot identify the
    status of most events with confidence, it can still correctly
    estimate the rates (Figure \ref{fig:analytic-rate-recovery}).}
\end{figure}

In Figure \ref{fig:analytic-rate-recovery}, we show the marginalized
posterior densities for the foreground and background rates (see
Eq.~\eqref{eq:rate-shape-posterior}).  Figure
\ref{fig:analytic-rate-foreground-probs} shows the posterior
foreground probability for each event marginalized over all other
events' types and the foreground and background rates.

\begin{figure}
  \includegraphics[width=\columnwidth]{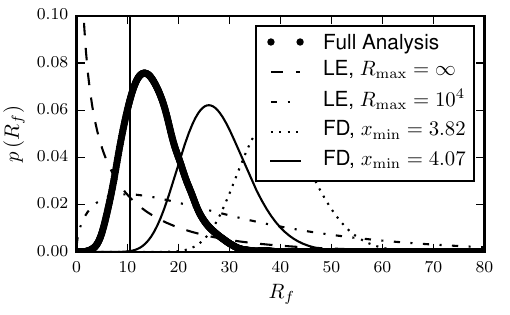}
  \caption{\label{fig:altmeth}Posteriors on foreground rate obtained
    using the method described in this paper, the loudest event
    statistic and the foreground dominated analysis for the data set
    from \S \ref{sec:analytic-GW-example}.  For the loudest event
    statistic, we present the posterior with and without an upper
    limit on the background rate, $R_b$; in both cases the rate
    posterior is significantly wider than the one obtained with the
    method described in this paper.  For the foreground dominated
    statistic, the limits $x_\mathrm{min} = 3.82$ and $x_\mathrm{min}
    = 4.07$ give likelihood ratios of $\hat{f}/\hat{b} = 0.5$ and
    $0.99$.  For this data set, the thresholds in fact include 19 and
    7 background events, respectively, so the corresponding rate
    estimates are significantly biased.  An ``omniscient'' threshold
    of $x_\mathrm{min} = 4.38$ would produce exactly 7 foreground and
    zero background events, and the resulting posterior is essentially
    indistinguishable from the curve for the full analysis.}
\end{figure}

We can compare these results to results obtained using the two
approximations described earlier, the loudest-event statistic and the
foreground-dominated statistic. The marginalized distribution for the
foreground rate using these alternatives are shown in
Figure~\ref{fig:altmeth}. In this case, the loudest event had
$x_N\simeq 9.47$. The loudest-event statistic depends on a
specification of the maximum, $R_{\rm max}$, for the background
rate. We show results for $R_{\rm max}=\infty$, i.e., the improper
prior, and $R_{\rm max}=10000$. The results for other reasonable
choices of $R_{\rm max} = 100, 1000, 100000$ etc. gave exactly the
same posterior, since $\hat{b}(x_N) R_{\rm max} \ll \hat{f}(x_N)$ for
all these choices and we are therefore in the regime where the
posterior is insensitive to $R_{\rm max}$. To apply the
foreground-dominated statistic we must specify a threshold above which
we assume all events are foreground. It is reasonable to do this based
on a specification for the relative probability of an event being
fore/background, $\hat{f}(x)/\hat{b}(x)=p_{\rm thresh}$. Setting
$p_{\rm thresh}=0.99$ gives $x_{\rm min}=4.07$ and there are $N=18$
(11 foreground and 7 background) events exceeding that
threshold. Setting $p_{\rm thresh}=0.5$ gives $x_{\rm min}=3.82$ and
there are $N=30$ (11 foreground and 19 background) events exceeding
that threshold.  Each of these thresholds gives a biased estimate of
the rate because there are background events still above threshold.
The ``omniscient'' threshold of $x_\mathrm{min} = 4.38$ produces $N=7$
(7 foreground and 0 background) events in this data set, and therefore
an unbiased estimate, but of course this threshold can only be
determined because we can examine the synthetic foreground and
background data samples. The threshold may seem obvious from a 
visual examination of Figure \ref{fig:analytic-rate-foreground-probs}; 
however, the construction of this figure relies on the application of the full
framework in the first place.  We show results for the first two choices of
$x_{\rm min}$ in Figure~\ref{fig:altmeth}; the omniscient choice
produces essentially the same posterior as our full analysis.

The loudest event statistic with the improper prior gives, as
expected, a poor approximation to the foreground rate. The peak is
more accurately located when a prior maximum rate is defined, but the
distribution is much wider than using the full analysis described here
in any case. This is to be expected as much of the information is
being thrown away. The foreground-dominated statistic gives a
reasonable approximation to the true foreground rate, and a
distribution that is essentially equal to the full analysis, for the
``omniscient'' choice of threshold value that excludes all background
data. For lower thresholds, even for a threshold where
$p_\mathrm{thresh} = 0.99$, it performs poorly since we are
approximating the foreground rate by the total foreground plus
background rate.  This indicates that, provided the threshold is
chosen appropriately, the foreground dominated statistic can perform
quite well at estimating the rate---but choosing this threshold
correctly is difficult. The fact that it reproduces the posterior from
the full analysis so well is indicative of the fact that most of the
information about the foreground comes from the loudest events. The
full analysis naturally incorporates inference about the background
rate $R_b$ along with the foreground rate and incorporates maximum
information from the data set and should therefore lead to narrower
posteriors in general.

\subsection{Gravitational Waves With Overlapping Templates}
\label{sec:gw-overlapping-template}

In \S \ref{sec:analytic-GW-example} we assumed that the overlap
between different templates in the template bank was negligible, so
the SNRs recovered by different templates are independent random
variables.  In fact, template banks are not constructed in this way
\citep[e.g.,][]{Owen:1998dk,Ajith:2008},
because signals could fall in the gaps between the non-overlapping
templates.  We can model this effect by assuming that a template bank
of $N$ actual templates will behave as if it had $N_\mathrm{eff}$
\emph{independent} templates.  Rather than pre-computing
$N_\mathrm{eff}$, we can fit for it as a shape parameter.  That is, we
assume that $\theta = \{N_\mathrm{eff}\}$ is a shape parameter for the
background cumulative distribution:
\begin{equation}
  \hat{B}\left(x, N_\mathrm{eff}\right) = \frac{\left( 1 + \erf\left(
    \frac{x}{\sqrt{2}} \right) \right)^{N_\mathrm{eff}} - \left( 1 +
    \erf\left( \frac{\xmin}{\sqrt{2}} \right)
    \right)^{N_\mathrm{eff}}}{2^{N_\mathrm{eff}} - \left( 1 +
    \erf\left( \frac{\xmin}{\sqrt{2}} \right) \right)^{N_\mathrm{eff}}
  }.
\end{equation}

\begin{figure}
  \includegraphics[width=\columnwidth]{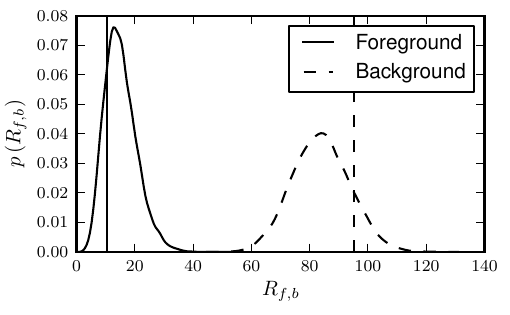}
  \caption{\label{fig:rates-nt} The foreground (solid lines) and
    background (dashed lines) rate posterior, marginalized over all
    flags and the $N_\mathrm{eff}$ parameter, for the gravitational
    wave template detection scenario with overlapping templates
    discussed in \S \ref{sec:gw-overlapping-template}.  The true
    values of the rates, $R_f = 10.4$ and $R_b=95.1$, are indicated
    with vertical lines.  The distributions are not significantly
    wider than those of Figure \ref{fig:analytic-rate-recovery}, in
    spite of the extra parameter.}
\end{figure}

\begin{figure}
  \includegraphics[width=\columnwidth]{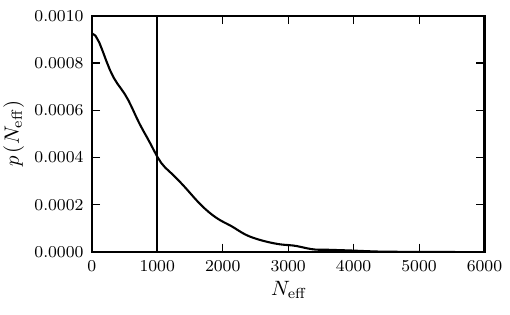}
  \caption{\label{fig:ntemplates} The posterior on the number of
    effective templates, $N_\mathrm{eff}$, for the model and data
    discussed in \S \ref{sec:gw-overlapping-template}, marginalized
    over all state flags and rates.  The true value, $N_\mathrm{eff} =
    1000$, is indicated by the vertical line.}
\end{figure}

Results from such an analysis appear in Figures \ref{fig:rates-nt} and
\ref{fig:ntemplates}.  We use the same parameters and data set as in
\S \ref{sec:analytic-GW-example}, with $x_\mathrm{min} = 3.5$, $R_f =
10.4$, $R_b = 95.1$, and $N_\mathrm{eff} = 1000$, but now allow
$N_\mathrm{eff}$ to be a parameter of the background distribution,
with a flat prior.  Both the rates and the number of effective
templates are recovered without significant loss of accuracy relative
to the fixed $N_\mathrm{eff}$ situation in \S
\ref{sec:analytic-GW-example}.

If we consider the two alternative methods, the loudest event and
foreground dominated statistics, and apply the same
foreground-dominated thresholds as before, we will recover the same
foreground distributions as are shown in Figure~\ref{fig:altmeth}.
This is because the parameter $N_{\rm eff}$ affects only the
background distribution, to which the foreground-dominated statistic
is insensitive, and in the loudest event case, after marginalization
over $N_{\rm eff}$ we find $\int_0^{N_{\rm max}} \hat{b}(x_N,N_{\rm
  eff}) {\rm d}N_{\rm eff} \ll 3 N_{\rm max}/R_{\rm max} \hat{f}(x_N,
N_{\rm eff})$ and so we are still in the foreground-dominated regime
in which the loudest event tells us nothing about the background.
Neither of these alternative methods can inform us about the value of
$N_{\rm eff}$, a property of the distribution of background events
identified by filtering with this template bank.  Moreover, the choice
of threshold value for the foreground-dominated statistic becomes
significantly more complicated in this case, since $p_\mathrm{thresh}$
now depends on $N_\mathrm{eff}$.

\subsection{Uncertainty in the Foreground and Background Distributions}
\label{sec:uncertain-dist}
The framework outlined above relies on the existence of models for the foreground, $\hat{f}(x,\theta)$, and background, $\hat{b}(x,\theta)$, distributions parameterized by a small number of model parameters, $\theta$. While in many situations simple analytic functions such as power laws will provide an adequate description, this will not always be the case. In the absence of a good analytic model, the space of the ranking statistic $x$ could be divided into bins and $\hat{f}(x)$ and $\hat{b}(x)$ are taken to be flat in each of these bins. The number of free parameters characterizing each of $\hat{f}$ and $\hat{b}$ is then the number of bins used. While such a framework is model free, the increase in model parameters will mean that more observed events will typically be required to achieve the same precision on the rates and foreground/background distributions.

In the context of gravitational wave experiments, additional information on the ranking statistic distributions for the foreground can be obtained using mock signal injections into the data, while distributions for the background can be estimated by analyzing time slides of data sets from different detectors relative to each other \cite[e.g.,][]{ihope}. This information can be readily incorporated in the current framework by assuming there is another set of $N_I$ events with ranking statistics $\{ w_i \}$, known to be drawn from the foreground distribution ($g_i=1$) and a set of $N_T$ events with ranking statistics $\{z_i\}$ known to be drawn from the background distribution ($g_i=0$). These events will typically not be drawn with the correct rate parameters, so they do not contribute to the estimates of $R_f$ and $R_b$, but they do contribute an extra factor
\be
\prod_{l=1}^{N_I} \hat{f}(w_l, \theta ) \prod_{m=1}^{N_T} \hat{b}(z_i, \theta) 
\ee
to the right hand sides of Eqs~(\ref{eq:posterior}) and (\ref{eq:rate-shape-posterior}). This approach provides a way to incorporate extra information into the analysis in order to simultaneously fit for the shape of the background and foreground as well as the rates. In the limit that there are many more events in the timeslide and injection data set, this will reduce to the analysis that was described above with fixed ranking statistic distributions $\hat{f}(x)$ and $\hat{b}(x)$ given by the injection and time slide data. We note that this analysis makes the assumption that the background distribution is the same in the time slide and real data and that the foreground distribution is the same between the injection and real data. The former assumption is probably reasonable, modulo correlations of non-gravitational-wave origin between data in different detectors, but the latter relies on knowledge of the relative the astrophysical rates of different events, which is more uncertain. These astrophysical uncertainties could be handled with a hybrid approach, in which injections are used to characterize the statistic distribution for sources of a particular type, while additional rate or shape parameters are introduced to characterize the variation in the astrophysical rate of mergers as a function of source type.

\subsection{Star Cluster Parameters With Background Contamination}
\label{sec:star-cluster}

Our final example concerns fitting for the location and shape
parameters of a cluster of stars observed on top of a stellar
background with a density gradient.  In this example, stars are either
members of the cluster (i.e.~foreground) or background contamination,
with a spatially varying density (i.e.~our rate functions are
two-dimensional).  Our method of analysis here is similar to that of
\citet{DeGennaro2009}, but here we marginalize over membership flags
and are simultaneously fitting foreground and background densities
(i.e.\ rates) \emph{and} cluster properties.  

We assume that a star cluster has a Plummer surface-density profile
\citep{Plummer1911,Aarseth1974},
\begin{equation}
  \label{eq:plummer-surface-density}
  \hat{f}(\vec{x}, \theta) = \frac{1}{\pi r_0^2 \left( 1 +
    \frac{\left| \vec{x} - \vec{x}_0 \right|^2}{r_0^2} \right)^2},
\end{equation}
where $\vec{x}_0$ is the location on the sky of the center of the
cluster, $r_0$ is a radial scale parameter, and $\vec{x} = \left( x, y
\right)$ is the position on the sky.  We assume a square observational
domain\footnote{The observational domain is not infinite, so the
  normalization of the cluster density in
  Eq.~\eqref{eq:plummer-surface-density} is not quite correct.  In our
  modeling we properly take this into account, but for simplicity here
  we ignore it.}, $\vec{x} \in [0,1]^2$, and a background that has a
density gradient at an arbitrary orientation with respect to the
observational axes:
\begin{equation}
  \hat{b}\left(\vec{x}, \theta\right) = 1 + \vec{\gamma} \cdot \left(
  \vec{x} - \vec{x}_{1/2} \right),
\end{equation}
where $\vec{\gamma}$ is the gradient, and $\vec{x}_{1/2} = [1/2, 1/2]$
is the centroid of the observational domain.  

\begin{figure}
  \includegraphics[width=\columnwidth]{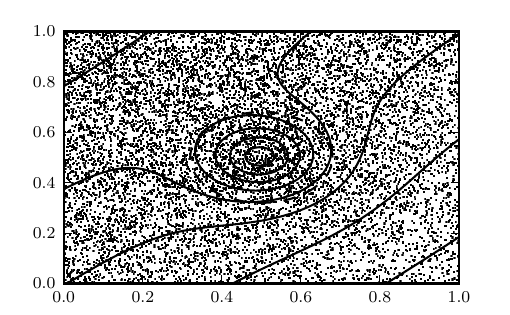}
  \caption{\label{fig:sky-density} Density contours and synthetic data
    for the example in \S~\ref{sec:star-cluster}.  The contours
    describe the true density profile with the parameters in
    Eq.~\eqref{eq:true-cluster-parameters}.  The points are the
    realization of this density profile used as synthetic data in
    \S~\ref{sec:star-cluster}; the dashed line encloses one Plummer
    scale radius about the true cluster center.  Because the peak
    cluster density is equal to the background density at the cluster
    center, the cluster is barely apparent to the eye.}
\end{figure}

We use simulated data drawn from our model with parameters
\begin{equation}
\label{eq:true-cluster-parameters}
\theta_0 \equiv \left\{ x_0, y_0, r_0, \gamma_x, \gamma_y \right\} =
\left\{ \frac{1}{2}, \frac{1}{2}, 0.18, -\frac{1}{2}, \frac{1}{2}
\right\},
\end{equation}
with $R_f = 1000$ and $R_b = 10000$.  For this set of parameters, the
average density of the background and the peak density of the cluster
are comparable; there are an order of magnitude more background stars
than cluster stars in the field.  Figure \ref{fig:sky-density} shows
the density of stars on the sky and the particular synthetic data set
used for this analysis.  Because the peak density of the cluster is
equal to the background density at the center of the domain, there is
no single star in the domain that is more likely to be a cluster
member than a background star (i.e.\ $\left \langle g_i \right \rangle
\lesssim 0.5$ for all stars); nevertheless, we will see that our
method provides good constraints on the cluster parameters.

To analyze our synthetic data set, we analytically marginalized over
the state flags (i.e.~cluster membership), using the likelihood in
Eq.~\eqref{eq:rate-shape-posterior}.  We did this to take advantage of
the \texttt{emcee} sampler of \citet{ForemanMackey2012}, which
requires all parameters to be in $\mathbb{R}$.  We applied a prior on
the shape parameters that is flat in $\vec{x}_0$ and $\vec{\gamma}$,
and an (approximately) Jeffreys prior on $r_0$,
\begin{equation}
  p\left( r_0 \right) = \frac{\sqrt{R_f}}{r_0}.
\end{equation}
(Note that this factor of $\sqrt{R_f}$ cancels with the Jeffreys prior
on the rate, $1/\sqrt{R_f}$; we have verified that the priors on these
parameters are irrelevant to our results, as would be expected from
the measurement of $\sim 1000$ foreground stars.)

\begin{figure}
  \includegraphics[width=\columnwidth]{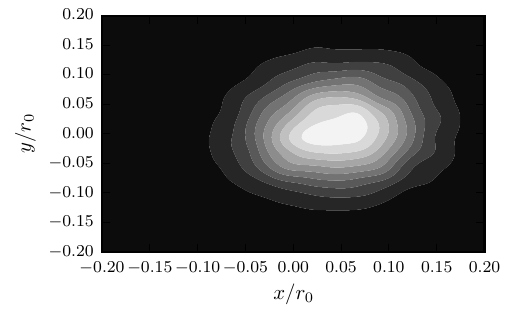}
  \caption{\label{fig:cluster-loc} Contours of the posterior
    probability distribution for the center of the cluster,
    $\vec{x}_0$, for the example from \S~\ref{sec:star-cluster}.  The
    center $(x,y) = \left(x_0, y_0\right)$ is determined to within
    about 5\% of the structural radius of the cluster, $r_0$ (see
    Eq.~\eqref{eq:true-cluster-parameters}).}
\end{figure}

\begin{figure}
  \includegraphics[width=\columnwidth]{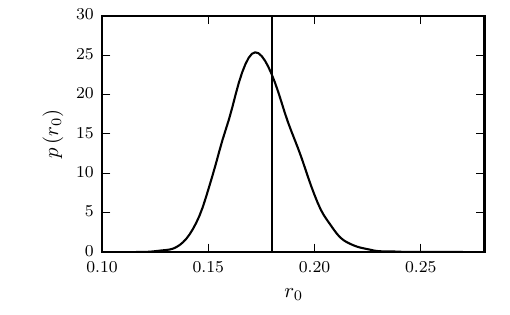}
  \caption{\label{fig:cluster-scale} Posterior density for the scale
    parameter for the cluster, $r_0$, for the example from
    \S~\ref{sec:star-cluster}.  The true value is indicated by the
    vertical line (see Eq.~\ref{eq:true-cluster-parameters}).}
\end{figure}

\begin{figure}
  \includegraphics[width=\columnwidth]{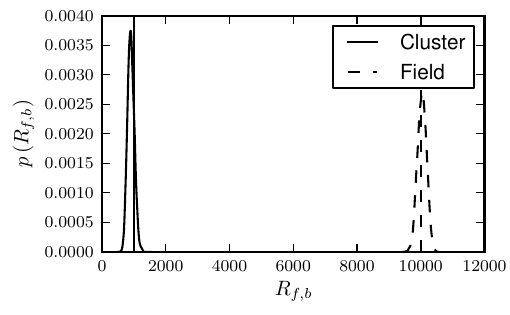}
  \caption{\label{fig:cluster-number} Posterior densities for the
    number of stars in the cluster ($R_f$) and in the field ($R_b$) in
    the example from \S~\ref{sec:star-cluster}.  Vertical lines
    indicate the true values (see
    Eq.~\eqref{eq:true-cluster-parameters}). }
\end{figure}

Figures \ref{fig:cluster-loc} and \ref{fig:cluster-scale} shows the
posteriors for the cluster location and scale parameters.  The center
of the cluster, $\vec{x}_0$, is localized to within about 5\% of the
cluster scale, and the cluster radius with a relative error of about
10\%.  In spite of the significant background, the cluster parameters
are recovered to a relative accuracy consistent with the expected
uncertainty from $N_\mathrm{eff} \simeq R_f = 1000$ measurements.
Figure \ref{fig:cluster-number} shows the posteriors inferred on the
cluster and background numbers, $R_f$ and $R_b$.

\section{Discussion}\label{sec:discussion}

In this paper, we have developed a Bayesian framework for rate
estimation when the data consists of a mixture of foreground and
background events.  We demonstrated the application of this framework
using several examples from gravitational-wave data analysis in the
presence of signatures of binary mergers and noise triggers, and
astronomical image analysis in the presence of several populations of
stars.  We showed that this framework is generally superior to both
the loudest-event statistic and the foreground-dominated
statistic.

Through most of this paper, we have assumed that the shape of the
foreground and background distributions is known, or at least can be
modeled with several additional parameters.  This is not necessarily
easy to do.  For example, in the case of gravitational-wave data
analysis, the shape of the foreground distribution of events may
depend on the details of a complex data-analysis pipeline as well as
the astrophysical source distribution, while the background event
distribution depends on data quality and may deviate significantly
from the simple Gaussian-noise behavior modeled in section
\ref{sec:GW-example}.  Several approaches have been developed to
accurately model both distributions, e.g., through the use of injected
signals \cite{ihope} or other methods \cite{CannonHannaKeppel:2012} to
model the foreground distribution.  However, this is a difficult
problem (e.g., because of the need to estimate the background at the
very tails of the distribution), and will require significant future
work.  In Section \ref{sec:uncertain-dist}, we discussed some of the
possible approaches when the shapes of the background and foreground
distributions cannot be confidently described by models with a few
adjustable parameters.

A further complication is that we have considered the rate of events
in the data as products of some analysis pipeline.  This rate may be
different from the physical rate of interest, such as the rate of
compact-binary mergers per unit time per unit volume which generate
gravitational waves, or the physical numbers of stars in the cluster
and field populations which produce the observed luminosities.  Again,
the conversion between the two will depend on the details of the
data-analysis algorithm and ranking statistic, including any selection
effects \cite{Messenger2012}, and would need to be determined on a
case-by-case basis.  See Ref.~\cite{Brady2008} for an example of such
conversion when the underlying framework is the loudest-event
statistic.

Furthermore, in a practical application there could be multiple
classes of events, not just foreground and background.  For example,
we are not necessarily interested in the rate of gravitational-wave
signals per se, but separately in the rate of signals from mergers of
binary neutron stars and binary black holes -- populations that may
sometimes be difficult to distinguish.  Our approach is readily
extendable to this particular complication, however.  Note that it is
symmetric with respect to foreground and background events (as
expected, since one physicist's background is another physicist's
foreground).  We could relabel foreground and background events into
other competing event classes, and further classes could be added in a
straightforward way.  However, the ability to distinguish classes
relies on different distributions of their statistics.  In general,
rankings may need to be extended to include other statistics in
addition to the signal ``loudness'' statistic in order to indicate
both event significance and the probability of event attribution to a
particular class.

\begin{acknowledgments}
  We thank Kipp Cannon, Thomas Dent, Chad Hanna, Drew Keppel, Richard
  O'Shaughnessy, David Hogg, and Ted von Hippel for discussions and
  suggestions about this manuscript.  IM and WMF acknowledge the
  hospitality of KITP, supported in part by the National Science
  Foundation under NSF Grant PHY11-25915.  CC's work was carried out
  at the Jet Propulsion Laboratory, California Institute of
  Technology, under contract to the National Aeronautics and Space
  Administration.  CC also gratefully acknowledges support from NSF
  grant PHY1068881.  JG's work is supported by the Royal Society.
  Copyright 2013.  All rights reserved.
\end{acknowledgments}

\bibliographystyle{apsrev4-1} \bibliography{many}

\end{document}